\documentclass[aps,twocolumn]{revtex4}
\usepackage{amssymb}
\usepackage{amsmath}
\usepackage{epsfig}
\begin{document}
%__________________________________________________________________________
%
%%\documentstyle[preprint,aps,amssymb]{revtex}  % PREPRINT format REVTEX 4.0
%\documentclass[draft,twocolumn,aps]{revtex4} % TWOCOLUMN format REVTEX 4.0
%%\documentstyle[aps]{revtex}           % GALLEY format REVTEX 4.0
%option: preprint

                           % this command makes pacs numbers print

%-----------------------------------------------------------------------------

%\title{\vskip -0.5in\hfill\hfil{\rm\normalsize Printed on \today}\vskip 0.4in

\title{Photovoltaic effect in a gated two-dimensional electron gas in magnetic field}                           

\author{Maria Lifshits$^{1,2}$ and Michel I. Dyakonov$^{1}$}  

\affiliation{$^{1}$Laboratoire de Physique Th\'eorique et Astroparticules, Universit\'e 
         Montpellier II, CNRS, France\\ $^{2}$ A.F. Ioffe Physico-Technical Institute, 
	 194021, St. Petersburg, Russia}

%\date{Received \quad}

%-----------------------------------------------------------------------------

\begin{abstract}
The photovoltaic effect induced by terahertz radiation in a gated two-dimensional electron gas 
in magnetic field is considered theoretically. It is assumed that the incoming radiation creates an 
ac voltage between the source and gate and that the gate length is long compared to the damping 
length of plasma waves. In the presence of pronounced Shubnikov-de Haas oscillations, an 
important  source of non-linearity is the oscillating dependence of the mobility on the ac gate 
voltage. This results in a photoresponse oscillating as a function of magnetic field, which is 
enhanced in the vicinity of the cyclotron resonance, in accordance with recent experiments. 
Another smooth component of the  photovoltage, unrelated to SdH oscillations,  has a maximum  
at cyclotron resonance. 


\end{abstract}

\maketitle

%-----------------------------------------------------------------------------

The two-dimensional gated electron gas in a Field Effect Transistor can be used 
for generation \cite{dyakonov1} and detection \cite{dyakonov2} of THz radiation, and both effects were 
demonstrated experimentally \cite{emission1, emission2, knap3, knap1, burke}. Concerning the 
detection, the idea 
is that the nonlinear properties of the electron fluid will lead to the rectification of the ac current 
induced in the transistor channel by the incoming radiation. As a result, a 
photoresponse in the form of a dc voltage between source and drain appears, which is proportional to the  
radiation intensity (photovoltaic effect). Obviously some asymmetry between the source and drain is 
needed to induce such a voltage. 

There may be various reasons of such asymmetry. One of them is  the difference in the source and 
drain boundary conditions. Another one  is the asymmetry in feeding the incoming radiation, which  
can be achieved either by using a special antenna, or by an asymmetric design of the source and 
drain contacts with respect to the gate contact. Thus the radiation may predominantly create an ac voltage 
between the source and the gate. Finally, the asymmetry can naturally arise if a dc current is passed 
between source and drain, creating a depletion of the electron density on the drain side of the channel.

The photoresponse can be either resonant, corresponding to the excitation of the descrete plasma 
oscillation modes in the channel, or non resonant, if the plasma oscillations are overdamped 
\cite{dyakonov2}. Both non-resonant \cite{knap3} and resonant \cite{knap1, burke} detection were 
demonstrated experimentally. A practically important case is that of a long gate, such that the plasma 
waves excited by the incoming radiation at the source cannot reach the drain side of the channel because 
their damping length is smaller than the source-drain distance. Within the hydrodynamic approach the 
following result for the photoinduced voltage was derived for this case \cite{dyakonov2}:
$$U= \frac{1} {4}\frac{U_a^2} {U_0}f(\omega), \quad  f(\omega)=1+\frac{2\omega\tau}{\sqrt{1+(\omega\tau )^2}}, 
\eqno{(1)}$$
where $\omega$ is the radiation frequency, $\tau$ is the momentum relaxation time, $U_a$ is the amplitude 
of the ac modulation of the gate-to-source voltage by the incoming radiation and $U_0$ is the static value 
of the gate-to-channel voltage swing, $U$, which is related to the electron density, $n$, in the channel 
by the plane capacitor formula:
$$ en= CU. \eqno{(2)}$$
Here, $e$ is the elementary charge, and $C$ is the gate-to-channel capacitance per unit area. Eq. (2)  
is applicable if the scale of the variation of the potential in the channel is large compared 
to the gate-to-channel separation.

Recently, the first experiments on the photovoltaic effect at terahertz frequencies in a gated high 
mobility two-dimensional electron gas in a magnetic field were performed \cite{CR1, CR2}. The 
main new results are: (i) the photoinduced dc drain-to-source voltage exhibits strong oscillations as a 
function of magnetic field, similar to the Shubnikov-de Haas (SdH) resistance oscillations, and 
(ii) the oscillation amplitude strongly increases in the vicinity of the cyclotron resonance.

In this Letter we consider theoretically the photovoltaic effect in a gated electron gas in a 
magnetic field assuming, as in Ref. \cite{dyakonov2}, that the incoming radiation creates an ac 
voltage between the source and the drain. Further, in accordance with the experimental conditions we 
assume that 1) the source-drain length, $L$, ($x$ direction) is greater than the plasma wave 
damping length, so that the plasma waves excited near the source do not reach the drain, and 2) 
the sample width, $W$, in the $y$ direction,  is much greater than $L$, see Fig. 1. The first assumption 
means that  the boundary conditions at the drain are irrelevant and, as far as plasma waves are 
concerned, the sample can be considered to be infinite in the $x$ direction. The second one implies a 
quasi-Corbino geometry (all variables depend on the $x$ coordinate only).

We explain the observed strongly oscillating photoresponse as being due to the non-linearity 
originating from the oscillating dependence of the mobility on the Fermi energy, and hence 
on the ac part of the gate voltage.
 
\begin{figure}
\epsfxsize=240pt {\epsffile{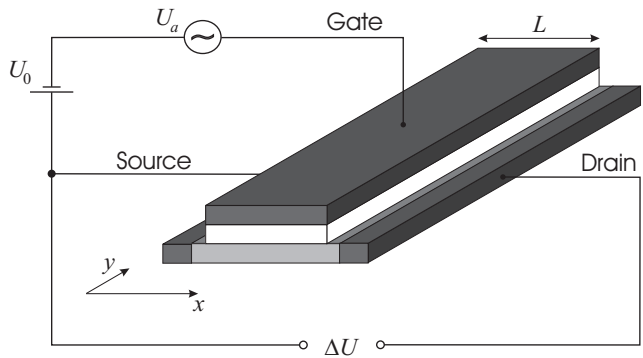}}
\caption{Assumed design and geometry. The THz radiation produces an ac voltage $U_a$ between the 
source and the gate inducing a dc source-drain voltage $\Delta U$. The gate width, $W$, is much 
larger than the gate length $L$ (quasi-Corbino geometry)}
\end{figure}

The photovoltaic effect is due to a radiation-induced force ${\bf G}$ driving the electron current.  
Without magnetic field, ${\bf G}$ is obviously directed in the $x$ direction and is compensated by 
the appearance of an electric field. In the presence of magnetic field the problem becomes more subtle, 
not only because in this case ${\bf G}$ has a $y$-component, but also because this radiation-induced 
force becomes {\it non-potential}: curl\,${\bf G}\neq 0.$  The non-potential part will drive an electric 
current along closed loops.

The significance of the cyclotron resonance for the photovoltaic effect is related to the 
well-known dispersion relation for plasma waves in a magnetic field \cite{quinn}. For gated 
two-dimensional electrons it reads:
$$\omega=\sqrt{\omega_{c}^2+s^2k^2},\eqno{(3)}$$
where $\omega_c$ is the cyclotron frequency, $s$ is the plasma wave velocity, and $k$ is the
wavevector. Thus, the plasma waves can propagate only if $\omega_c<\omega$. In the opposite case 
the wavevector becomes imaginary, so that the plasma oscillations rapidly decay away from the source. 
The change of regime when the magnetic field is driven through its resonant value will manifest 
itself in the photoresponse.

Following Refs. \cite{dyakonov1, dyakonov2} and other theoretical work, we will use the hydrodynamical 
approach because, like the Drude equation, it provides a relatively simple description, compared to the 
full kinetic theory. However it should be understood that at low temperatures, at which the experiments 
\cite{CR1, CR2} were done, this approach strictly speaking is not justified, because the collisions 
between electrons are strongly suppressed by the Pauli principle. Nevertheless, the qualitative physical 
results derived from the kinetic equation and from the hydrodynamic equations are usually similar, e.g.  
the properties of plasma waves are identical in both approaches, provided that the plasma wave velocity 
$s$ is greater than the Fermi velocity, so that the Landau damping can be neglected \cite{dmitriev}. For 
this reason, we leave the much more complicated approach based on the kinetic equation for future studies.

The electrons in a gated 2D channel can be described by the following equations:
$$\frac{\partial {\bf v}}{\partial t} +({\bf v}\cdot \nabla){\bf v}=- \frac{e}{m}\nabla U + 
\frac{e}{mc}{\bf B} \times {\bf v} - \gamma {\bf v}, \eqno{(4)}$$
$$\frac{\partial U}{\partial t} + {\rm div}(U{\bf v})=0, \eqno{(5)}$$
where ${\bf v}$ is the electron drift velocity, ${\bf B}$ is the magnetic field along the 
$z$ direction, $m$ is the electron effective mass, and $\gamma=1/\tau$. The parameter 
$\gamma$ is an oscillating function of the electron concentration (or gate voltage) 
and magnetic field, which results in the SdH oscillations.

Eq. (4) is the Euler equation, taking account of the Lorentz force and damping due to collisions. 
It differs from the conventional Drude equation only by the convective term $({\bf v}\cdot \nabla)
{\bf v}$. Equation (5) is the continuity equation rewritten with the use of Eq. (2).

The boundary condition at the source ($x=0$) is:
$$U(0,t)=U_0 +U_a\cos \omega t,  \eqno{(6)}$$
where $\omega$ is the frequency of the incoming radiation, and $U_a$ is the amplitude of the 
radiation-induced modulation of the gate-to-source voltage. For a long sample, the boundary 
condition at the drain is 
$${\bf v}\rightarrow 0, \quad U\rightarrow U_0 {\rm \;\;for\;\;} {x\rightarrow \infty}.\eqno{(7)}$$

We will search for the solution of Eqs. (4) and (5) as an expansion in powers of $U_a$:
$${\bf v} = {\bf v}_1 + {\bf v}_2 , \quad U = U_0  +U_1 + U_2 . \eqno{(8)}$$
Here ${\bf v}_1$ and  $U_1$ are the ac components proportional to $U_a$, which can be found by 
linearizing Eqs. (4, 5), ${\bf v}_2$ and $U_2$ are the dc components, proportional to $U_{a}^2$ 
(we are not interested in the second harmonic terms $\sim U_{a}^2$). 
It is convenient to introduce  $u=eU/m$, $u_a=eU_a/m$, and the plasma wave velocity 
in the absence of magnetic field $s =u_{0}^{1/2}= (eU_0/m)^{1/2}$ \cite{dyakonov1}.  

To the first order in $U_a$, we obtain:
$$\frac{\partial v_{1x}}{\partial t} + \frac{\partial u_1}{\partial x} + \omega_c v_{1y} + 
\gamma v_{1x}=0 , \eqno{(9)}$$
$$\frac{\partial v_{1y}}{\partial t} - \omega_c v_{1x} + \gamma v_{1y}=0 , \eqno{(10)}$$
$$\frac{\partial u_{1}}{\partial t} + s^2\frac{\partial v_{1x}}{\partial x}=0, \eqno{(11)}$$
where $\omega_c=eB/mc$ is the cyclotron frequency. The boundary conditions follow from Eqs. (6, 7):
$ u_1(0,t)=u_a\cos(\omega t)$ and $ u_1(\infty,t)=0, {\bf v}_1(\infty,t)=0$.

Searching for the solutions $\sim\exp (ikx - i\omega t)$, 
we obtain the dispersion equation for the plasma waves:
$$\frac {s^2}{\omega^2} k^2= 1+i\alpha - \frac{\beta^2}{1+i\alpha},  \eqno{(12)}$$
where $\alpha = (\omega\tau)^{-1}$ and $\beta=\omega_c/\omega$ is the magnetic field in units of its 
resonant value for a given $\omega$. To ensure the boundary condition at $x\rightarrow\infty$ the root 
with a positive imaginary part of $k$ should be chosen. If damping is neglected ($\alpha=0$), this 
equation reduces to Eq. (3). The explicite expressions for $u_1$, $v_{1x}$, and $v_{1y}$ are easily 
obtained from Eqs. (9-11). 

In the second order in $U_a$, we find
$$\frac{du_2}{d x} + \omega_c v_{2y} + \gamma v_{2x}+\langle v_{1x}\frac {\partial v_{1x}}{\partial x}
\rangle +\gamma'\langle u_1v_{1x}\rangle =0,\eqno{(13)}$$ 
$$ -\omega_c v_{2x} + \gamma v_{2y}+\langle v_{1x}\frac {\partial v_{1y}}{\partial x}\rangle+\gamma' 
\langle u_1v_{1y}\rangle =0, \eqno{(14)}$$  
$$\frac{dj_x}{d x}=0, \qquad j_x=v_{2x} + \frac {1}{u_0 }\langle u_1 v_{1x}\rangle, \eqno{(15)}$$ 
where the angular brackets denote the time averaging over the period $2 \pi/\omega$. Here we have expanded 
the function $\gamma(u)$ to the first order in $u_1$. The quantities $\gamma$ and $\gamma'=d\gamma/du$ should 
be taken at $u=u_0$. The boundary conditions for Eqs. (13-15) are: $u_2(0)=0$, $v_{2x}(\infty)=v_{2y}(\infty)=0$.

From Eq. (15) we derive the obvious fact that $j_x=0$ ($j_x$ differs from the $x$ component of the true 
current density only by a factor $en$). Using this, and introducing the $y$ component of the current, 
$j_y$, by a relation similar to Eq. (15), we can rewrite Eqs. (13, 14) as follows: 
$$\omega_c j_y  = G_x(x)-\frac{du_2}{d x}, \qquad  \gamma j_y = G_y(x),\eqno{(16)}$$ 
where the additional driving force {\bf G} induced by the incoming radiation is given by:
$$G_x = \left (\frac{\gamma}{u_0}-\gamma'\right) \langle u_1v_{1x}\rangle +\frac {\omega_c}{u_0}\langle 
u_1v_{1y}\rangle - \langle v_{1x}\frac {\partial v_{1x}}{\partial x}\rangle ,\eqno{(17)}$$
$$G_y = \left (\frac {\gamma}{u_0} - \gamma'\right )\langle u_1v_{1y}\rangle -\frac {\omega_c}{u_0}\langle 
u_1v_{1x}\rangle - \langle v_{1x}\frac {\partial v_{1y}}{\partial x}\rangle.\eqno{(18)}$$
Both $G_x$ and $G_y$ depend on $x$ as $\exp (-2k''x)$, where $k''$ is the imaginary part 
of the wavevector defined by Eq. (12), reflecting the decay of the plasma wave intensity away from the source.
Thus curl${\bf G}\neq 0$.

One could solve Eqs. (16) to obtain the photoinduced voltage $\Delta u=\int^\infty_0
[G_x-(\omega_c/\gamma)G_y]\,dx$ and this would be the correct result for the {\it true} Corbino geometry, 
where the current $j_y$ can freely circulate around the ring. However, we believe that this is not 
correct for a finite strip, even if $W>>L$, because in this case the current $j_y$
induced by the non-potential part of the driving force, $G_y(x)$, obviously must return back somewhere,
forming closed loops \cite{remark}. How exactly this will happen, is not quite clear. In our model, the
current loops are likely to close through the source contact, however in reality the oppositely directed 
$y$-current will probably flow in the ungated part of the channel adjacent to this contact. Anyway, 
since the current $j_y$ integrated over $x$ must be zero (except near the extremities), we believe that 
the correct way is to integrate the first of equations (16) taking this into account,  and to ignore the 
second one, which is not applicable beyond the gated part of the channel. The integration interval should 
be expanded to include the region where the current lines return backwards. 

So far, we have no rigorous proof that this idea is correct, however we have checked that both methods 
give similar qualitative results (but differ in the exact form of the magnetic field dependence of the 
photovoltage).

As described above, we obtain $\Delta u=u_2(\infty)$:
$$\Delta u = \int^\infty_0G_x(x)\,dx. \eqno{(19)}$$
Using Eqs. (17, 19) we finally calculate the dc photovoltage $\Delta U= m \Delta u/e$, 
between drain and source induced by the incoming radiation: 
$$\Delta U = \frac{1}{4}\frac{U_a^2}{U_0}\left[f(\beta)-\frac {d\gamma}{dn}
\frac {n}{\gamma}\,g(\beta)\right]. \eqno{(20)}$$ 
Here we have separated the photoreponse in a smooth part and in an oscillating part. The second one, 
proportional to $d\gamma/dn$, is an oscillating function of gate voltage or magnetic field 
$\sim d\rho_{xx}/dn$, where $\rho_{xx}$ is the longitudinal resistivity of the gated electron gas.

Note, that even if the amplitude of the SdH oscillations is small, the parameter 
$|d\rho_{xx}/dn|(n/\rho_{xx})$ can be large, so that the oscillating contribution may dominate.
\begin{figure}
\epsfxsize=240pt {\epsffile{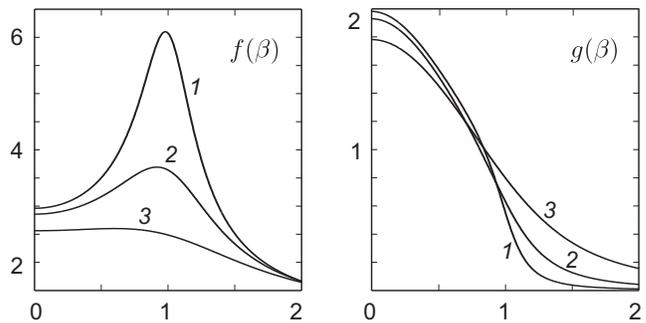}}
\caption{The functions $f(\beta)$ (left)  and $g(\beta)$ (right) describing respectively the smooth part 
and the envelope for the oscillating part of the photovoltage. The values of the parameter 
$\alpha=(\omega\tau)^{-1}$: {\it 1} - 0.2, {\it 2} - 0.4, {\it 3} - 0.8} 
\end{figure}

The frequency and magnetic field dependences of the photovoltage are described by the functions 
$f(\beta)$ and $g(\beta)$, which are given by the following formulas \cite{remark1}:
$$f(\beta)= 1+\frac {1+F}{\sqrt{\alpha^2+F^2}} ,\eqno{(21)}$$
$$g(\beta)= \frac{1+F}{2} \left ( 1+\frac {F}{\sqrt{\alpha^2+F^2}}\right ),\eqno{(22)}$$
where $F$ depends only on the ratio $\beta=\omega_c/\omega$ and the dimensionless parameter 
$\alpha=(\omega\tau)^{-1}$: 
$$F = \frac {1+\alpha^2 - \beta^2} {1+\alpha^2 + \beta^2}. \eqno{(23)} $$
In the absence of magnetic field, $\beta=0$, $F=1$, and Eq. (21) reduces to Eq. (1).

Figure 2 shows the behavior of the functions $f(\beta)$ and $g(\beta)$ for several values of the 
parameter $\alpha$. One can see that for small values of $\alpha$ (or large $\omega\tau$) the smooth part 
displays the cyclotron resonance with the unusual lineshape $f(\beta)\sim [(1-\beta)^2+\alpha^2]^{-1/2}$. 
The envelope for the oscillating part, $g(\beta)$ exhibits a fast decay beyond the cyclotron resonance 
($\beta>1$), confining the oscillations of the photovoltage to the region $\beta \sim 1$.

To display the oscillating contribution, we take the parameter $\gamma$ in the conventional form \cite{cole}, 
which is valid when the SdH oscillations are small:
$$\gamma = \gamma_0\left[1-4\frac {\chi}{\sinh \chi} \exp \left(-\frac {\pi}{\omega_c \tau_q}\right)\cos \left 
(\frac {2\pi E_F}{\hbar \omega_c}\right)\right], \eqno{(24)}$$
where $\chi =2\pi^2kT/\hbar \omega_c$, $\tau_q$ is the ``quantum" relaxation time, and $E_F$ is the Fermi 
energy, which is proportional to the electron concentration $n$, and hence to the gate voltage swing 
$U$. 
\begin{figure}
\epsfxsize=240pt {\epsffile{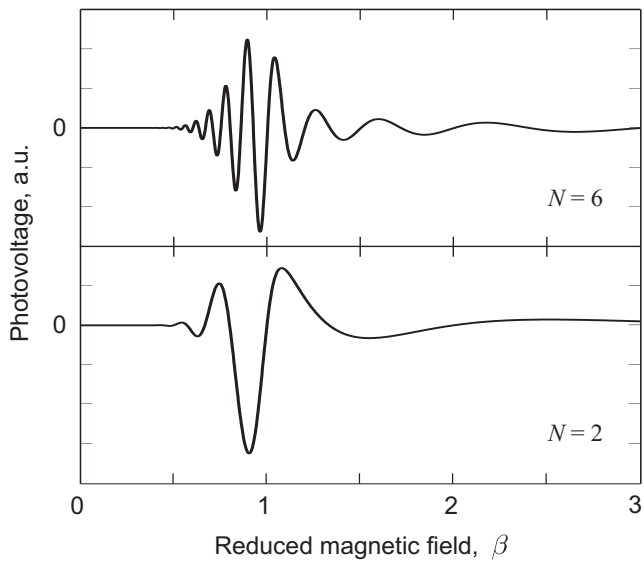}}
\caption{Magnetic field dependence of the oscillating part of the photovoltage for $\alpha=0.1$, 
$\omega\tau_q=0.5$ for two values of $N=E_F/\hbar \omega$. The vertical scale for the lower trace is 
expanded 4 times with respect to the upper trace. $\beta=\omega_c/\omega$}
\end{figure}

We introduce the parameter $N=E_F/\hbar \omega$, which is the number of Landau levels 
below the Fermi level at cyclotron resonance. Figure 3 presents the oscillating part of 
the photovoltage [the function $-(d\gamma/dn)(n/\gamma)g(\beta)$] for $\alpha=0.1$, $\chi =0.7$ 
(corresponding to $T=4$K, $\omega = 2\pi\cdot 2.5\,$THz), and $\omega \tau_q=0.5$, for two values of $N$.

In spite of the crudeness of our model, which does not account for various features of the experimental 
situation (the unavoidable presence of ungated parts of the channel, etc), our results show a good 
qualitative agreement with the recent experimental findings \cite{CR2}.

In summary, we have calculated the photovoltage induced in a gated electron gas by 
THz radiation in the presence of the magnetic field. As a function of magnetic field, 
the photoresponse contains a smoothly varying part and an oscillating part proportional to the derivative of 
the SdH oscillations with respect to the gate voltage. The smooth part shows an enhancement  
in the vicinity of the cyclotron resonance.

We appreciate numerous helpful discussions with Wojciech Knap, Nina Dyakonova, Maciej Sakovicz, 
St\`ephane Boubanga-Tombet, and Sergei Rumyantsev.


\begin{thebibliography}{}

\bibitem{dyakonov1} M. Dyakonov and M. Shur, Phys.  Rev. Lett. {\bf 71}, 2465 (1993)

\bibitem{dyakonov2} M. Dyakonov and M. Shur, IEEE Transactions on Electron Devices, 
{\bf 43}, 380 (1996)

\bibitem{emission1} W. Knap, J. Lusakowski, T. Parenty, S. Bollaert, A. Cappy and M.S. Shur,
Appl. Phys. Lett. {\bf 84}, 2331 (2004)		

\bibitem{emission2} N. Dyakonova, A. El Fatimy, J. Lusakowski, W. Knap, M.I. Dyakonov, 
M.-A. Poisson, E. Morvan, S. Bollaert, A. Shchepetov, Y. Roelens, Ch. Gaquiere, D. Theron, 
and A. Cappy, Appl. Phys. Lett. {\bf 88}, 141906 (2006)

\bibitem{knap3} W. Knap, V. Kachorovskii, Y. Deng, S. Rumyantsev, J.-Q. Lu, R. Gaska, M.S. Shur, 
G. Simin, X. Hu, M. Asif Khan, C.A. Saylor and L.C. Brunel, J. Appl. Phys. {\bf 91}, 9346 (2002)

\bibitem{knap1} W. Knap, Y. Deng, S. Rumyantsev, J-Q. Lu, M.S. Shur, C.A. Saylor, and L.C. Brunel,
Appl. Phys. Lett.  {\bf 43}, 3434 (2002)

\bibitem{burke} S. Kang, P.J. Burke, L.N. Pfeiffer, and K.W. West, Appl. Phys. Lett. {\bf 89}, 
213512 (2006)
	
\bibitem{CR1} M. Sakowicz et al, Int. J. High Speed Electron. Syst., to be published; 
M. Sakowicz et al,  Int. J. Mod. Phys. B, to be published

\bibitem{CR2} S. Boubanga-Tombet et al, Appl. Phys. Lett. to be published
	
\bibitem{quinn} K.W. Chiu and J.J. Quinn, Phys. Rev. B {\bf 9}, 4724 (1974)

\bibitem{dmitriev} A.P. Dmitriev, V.Yu. Kachorovskii, and M.S. Shur, Appl. Phys. Lett. {\bf 79}, 
922 (2001)

\bibitem{remark} In a Hall transport experiment there is no significant difference between the true 
Corbino geometry and the quasi-Corbino case of a finite strip with $W>>L$. The current $j_y$ exists 
everywhere, except the extremities of the sample at $y=\pm W/2$, where the current lines exit and enter
the left and right contacts respectively. In our case, the current lines must form closed loops, which 
most probably will pass through the source contact, or the adjacent to this contact ungated part of the 
channel

\bibitem{remark1} Similar results can be obtained within the Drude theory (neglecting the convective term 
$({\bf v}\cdot \nabla){\bf v}$). The oscillating part remains the same, while Eq. (21) aquires an
additional factor $1/2$ in the second term, which does not modify the qualitative behavior of $f(\beta)$

\bibitem{cole} P.T. Coleridge, R. Stoner, and R. Fletcher, Phys. Rev. B {\bf 39}, 1120 (1989)


\end{thebibliography}
\end{document}